\documentclass[runningheads]{llncs}
\usepackage{graphicx}
\usepackage{cite}
\usepackage{hyperref}
\usepackage{amsmath} 
\usepackage{amssymb}
\usepackage{bm}
\usepackage{booktabs}
\usepackage{makecell}
\usepackage{array,color}
\usepackage{array}
\usepackage[table]{xcolor}
\usepackage{multirow}

\usepackage[misc]{ifsym}

\begin{document}
\bibliographystyle{splncs04}

\title{FDA: Feature Decomposition and Aggregation for Robust Airway Segmentation}
\titlerunning{FDA: Feature Decomposition and Aggregation}
%

\author{Minghui Zhang\inst{1} \and 
Xin Yu\inst{3} \and 
Hanxiao Zhang\inst{1}  \and 
Hao Zheng\inst{1} \and
Weihao Yu\inst{1} \and
Hong Pan\inst{2} \and
Xiangran Cai\inst{3} \and
Yun Gu\inst{1,4}\textsuperscript{(\Letter)}
}


%
\authorrunning{M. Zhang et al.}
%
\institute{Institute of Medical Robotics, 
Shanghai Jiao Tong University, Shanghai, China
\email{
    geron762@sjtu.edu.cn}\\
\and
Department of Computer Science and Software Engineering, 
Swinburne University of Technology, Victoria, Australia \\
\and
Medical Image Centre, The First Affiliated Hospital of Jinan 
University, Guangzhou, China\\
\and
Shanghai Center for Brain Science and Brain-Inspired Technology, Shanghai, China
}
\maketitle              

\begin{abstract}
3D Convolutional Neural Networks (CNNs) 
have been widely adopted for airway segmentation. 
The performance of 3D CNNs is 
greatly influenced by the dataset while the public airway datasets 
are mainly clean CT scans with 
coarse annotation, thus difficult to be generalized to noisy CT scans (e.g. COVID-19 CT scans). 
In this work, we proposed a new dual-stream network to address the variability between the 
clean domain and noisy domain, which utilizes the clean CT scans and 
a small amount of labeled noisy CT scans for airway segmentation. 
We designed two different encoders to extract the transferable clean features and 
the unique noisy features separately, followed by two independent decoders. 
Further on, the transferable features are refined 
by the channel-wise feature recalibration and Signed Distance Map (SDM) regression. 
The feature recalibration module emphasizes critical features and the SDM
pays more attention to the bronchi, which is beneficial to extracting the transferable topological 
features robust to the coarse labels. Extensive experimental 
results demonstrated the obvious improvement brought by our proposed method. Compared to 
other state-of-the-art transfer learning methods, our method accurately 
segmented more bronchi in the noisy CT scans.
\end{abstract}

\section{Introduction} 
The novel coronavirus 2019 (COVID-19) has turned into a pandemic, infecting humans all 
over the world. To relieve the burden of clinicians, many researchers take the 
advantage of deep learning methods for automated COVID-19 diagnosis and infection 
measurement from imaging data (e.g., CT scans, Chest X-ray). Current studies mainly 
focus on designing a discriminative or robust model to distinguish COVID-19 from 
other patients with pneumonia\cite{wang2020prior,ouyang2020dual}, 
lesion localization\cite{wang2020weakly}, and segmentation\cite{wang2020noise}. In this work, 
we tackle another challenging problem, airway segmentation of COVID-19 CT scans. 
The accurate segmentation enables the quantitative measurements 
of airway dimensions and wall thickness which can reveal the abnormality
of patients with pulmonary disease and 
the extraction of patient-specific airway model from CT image is required 
for navigation in bronchoscopic-assisted surgery. It helps the sputum suction for 
novel COVID-19 patients.
\begin{figure}[!t]
    \centering
    \includegraphics[width=1.0\linewidth]{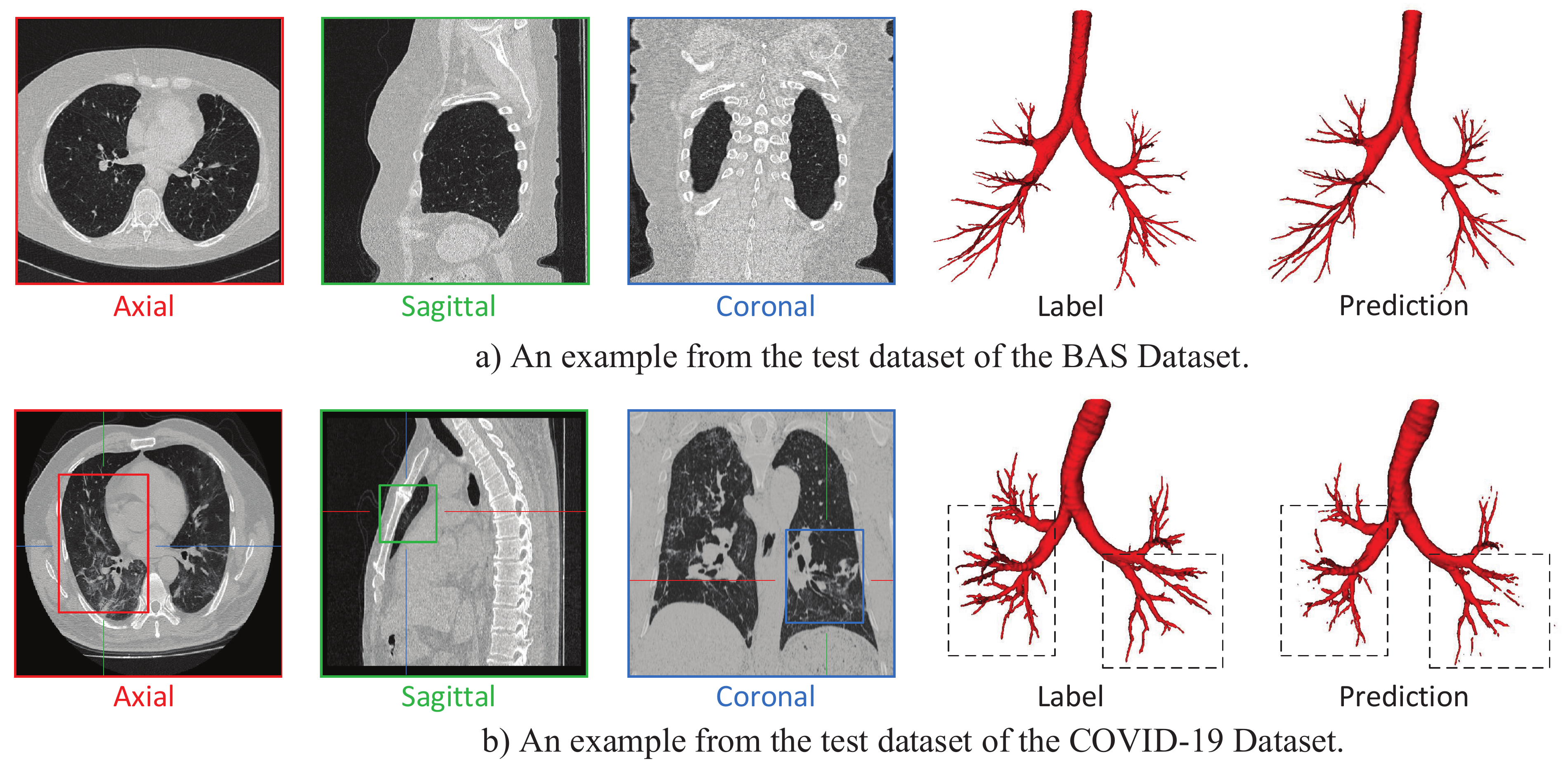}
    \caption{Different patterns exist between the BAS dataset and 
    the COVID-19 dataset. One well-trained segmentation model in the clean domain leads to 
    low accuracy when testing in the noisy domain.} \label{fig1}
\end{figure}

However, due to the fine-grained pulmonary airway structure, 
manual annotation is time-consuming, error-prone, and 
highly relies on the expertise of clinicians. Moreover, COVID-19 CT scans 
share ground-glass opacities in the early stage and pulmonary consolidation 
in the late stage\cite{chung2020ct} that adds additional difficulty for annotation. 
Even though the fully convolutional networks (FCNs) could automatically segment the airway, 
there remain the following challenges. First, FCNs are data-driven methods, while 
there are few public airway datasets with annotation and the data size is also limited. 
The public airway datasets, including EXACT'09 dataset\cite{lo2012extraction} and 
the Binary Airway Segmentation (BAS) dataset\cite{qin2019airwaynet}, 
focus on the cases with the abnormality of airway structures mainly caused by 
chronic obstructive pulmonary disease (COPD). 
These cases are relatively clean and we term their distribution as \textbf{Clean Domain}, on the contrary, 
we term the distribution of COVID-19 CT scans as \textbf{Noisy Domain}. Fig.~\ref{fig1} shows that 
fully convolutional networks (FCNs) methods\cite{juarez2018automatic,qin2019airwaynet} 
trained on the clean domain cannot be perfectly generalized to the noisy domain. Although this challenge can
be addressed via the collection and labeling of new cases, it is impractical for
novel diseases, e.g. COVID-19, which cannot guarantee the scale of datasets and
the quality of annotation. 
Second, transfer learning methods 
(e,g. domain adaptation\cite{rozantsev2018beyond,chen2020unsupervised}, 
feature alignment\cite{chen2020unsupervised,sun2016deep}) can improve the performance on 
target domains by transferring the knowledge contained in source domains or learning domain-invariant 
features. However, these methods are inadequate to apply in our scenario because this target noisy domain 
contains specific features (e.g. patterns of shadow patches) which cannot be learned from the source domain. 
Third, the annotation of the airway is extremely hard as 
they are elongated fine structures with plentiful 
peripheral bronchi of quite different sizes and orientations. 
The annotation in the EXACT'09 dataset\cite{lo2012extraction} and 
the BAS dataset\cite{qin2019airwaynet} are overall coarse and unsatisfactory. 
However, the deep learning methods are intended to fit the coarse labels, 
and thereby they are difficult to learn the robust features for airway representation. 
 
To alleviate such challenges, we propose a dual-stream network to extract 
the robust and transferable features from the clean CT scans (clean domain)
and a few labeled COVID-19 CT scans (noisy domain). Our contributions are threefold:
\begin{itemize}
    \item We hypothesize that the COVID-19 CT scans own the general features 
    and specific features for airway segmentation. The general features
    (e.g. the topological structure) are likely to learn from the other clean CT scans, 
    while the specific features (e.g. patterns of shadow patches) should be extracted 
    independently. Therefore, we designed a dual-stream network, 
    which takes both the clean 
    CT scans and a few labeled COVID-19 CT scans as input to synergistically 
    learn general features and independently 
    learn specific features for airway segmentation. 
    \item We introduce the feature calibration module and the Signed Distance Map (SDM) 
    for the clean CT scans with coarse labels, and through this way, 
    robust features can be obtained for the extraction 
    of general features.
    \item With extensive experiments on the clean CT scans and the COVID-19 CT scans, 
    our method revealed the superiority in the extraction of transferable and robust features 
    and achieved improvement compared to other methods under the evaluation of tree length 
    detected rate and the branch detected rate.
\end{itemize}


\section{Method}
\begin{figure}[!t]
    \centering
    \includegraphics[width=1.0\linewidth]{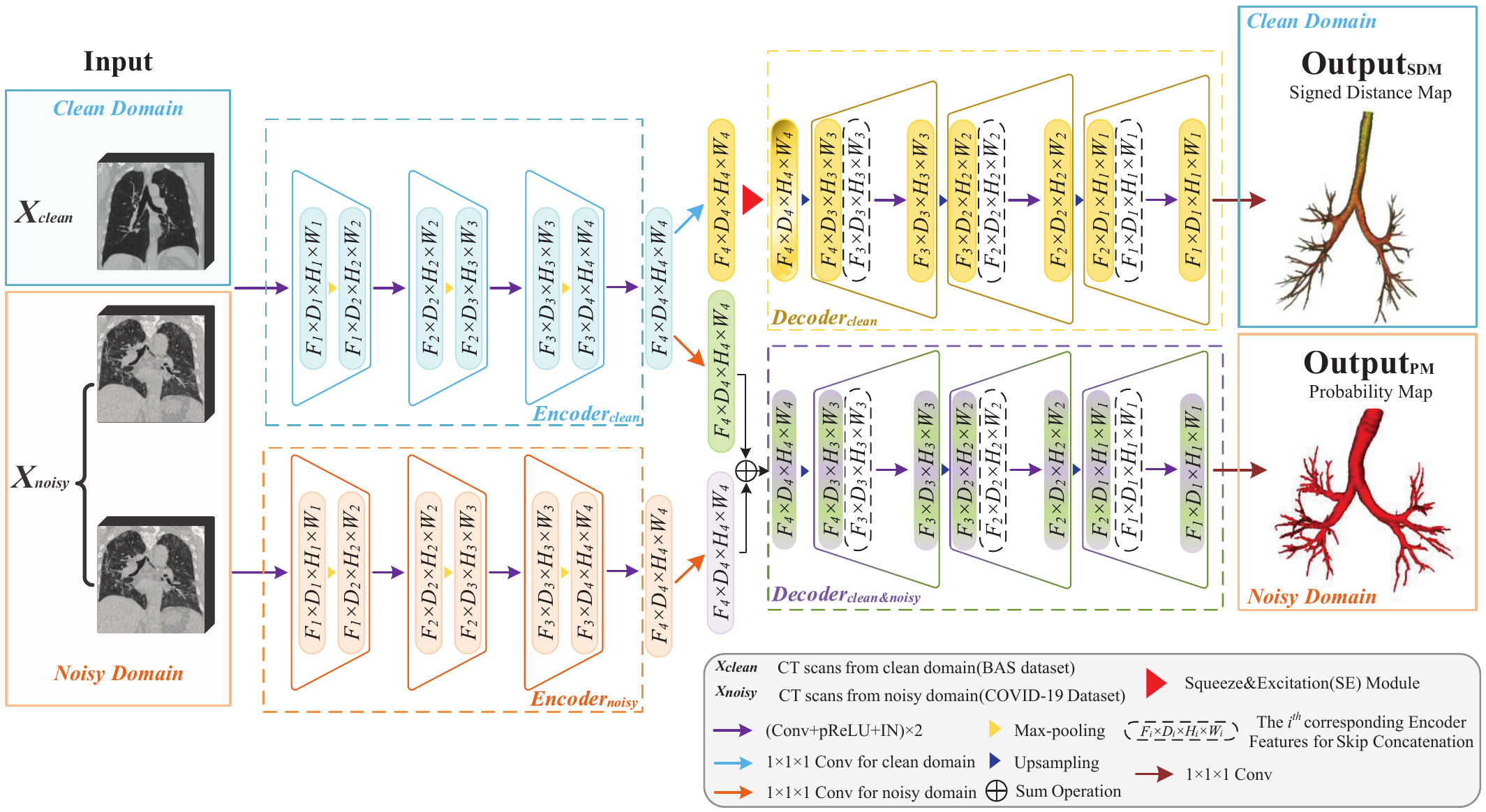}
    \caption{Detailed structure and workflow of the proposed dual-stream network.
    $Encoder_{clean}$ aims to synergistically 
    learn transferable features from both $X_{clean}$ and 
    $X_{noisy}$. $Encoder_{noisy}$ extracts specific features of $X_{noisy}$. 
    The encoded features of $X_{clean}$ are fed into $Decoder_{clean}$, where SE and 
    SDM modules refine the features.  
    For $X_{noisy}$, the decomposed features are then aggregated again via 
    the channel-wise summation operation and fed into $Decoder_{clean\&noisy}$.
      }\label{fig2}
\end{figure}

A new dual-stream network is proposed, which simultaneously processes 
the clean CT scans and a few noisy COVID-19 CT scans 
to learn robust and transferable features for airway segmentation. 
In this section, we detail the architecture of the 
proposed dual-stream network, which is illustrated in Fig.~\ref{fig2}.
\subsection{Learning Transferable Features} 
COVID-19 CT scans share a similar airway topological 
structure with the clean CT scans, 
meanwhile introduce unique patterns, e.g., multifocal patchy 
shadowing and ground-glass opacities, which are not common in clean CT scans. 
Since the number of clean CT scans is relatively large and the airway 
structure is also clearer, we aim to adapt the knowledge from clean CT 
scans to improve the airway segmentation of COVID-19 CT scans. Therefore, 
a dual-stream network is designed to synergistically learn transferable 
features from both COVID-19 CT scans and clean CT scans, and independently 
learn specific features only from COVID-19 CT scans. As illustrated in Fig.\ref{fig2}, 
let $\bm{X}_{clean}$ denote the input of sub-volume CT scans 
from the clean CT scans, $\bm{X}_{noisy}$ from the noisy COVID-19 CT scans. 
$Encoder_{clean}$ and $Encoder_{noisy}$ are encoder blocks for feature extraction. 
$Decoder_{clean}$ and $Decoder_{clean\&noisy}$ are decoder blocks to 
generate the segmentation results based on the features from encoders. 
The $Output_{SDM}$ represents the output of clean CT scans, and the $Output_{PM}$ 
represents the output of COVID-19 CT scans, they can be briefly defined as follows: 
\begin{eqnarray*}\label{eq1}
\begin{aligned}
    &{\rm Output_{SDM}} = Decoder_{clean}(Encoder_{clean}(\bm{X}_{clean})) \\
    &{\rm Output_{PM}} = Decoder_{clean\&noisy}(Encoder_{clean}(\bm{X}_{noisy}) + Encoder_{noisy}(\bm{X}_{noisy}))
\end{aligned}
\end{eqnarray*}
where we omit the detail of $1\times1\times1$ convolution and 
the Squeeze\&Excitation (SE) module for a straightforward 
explanation of the overall workflow. In this case, the features 
of $X_{noisy}$ are decomposed into two parts: $Encoder_{clean}$ aims 
to extract high-level, semantic and transferable features from both 
clean CT scans and COVID-19 CT scans; $Encoder_{noisy}$ is designed 
to obtain the specific features which belongs to the COVID-19 samples. 
The features of clean CT scans extracted by $Encoder_{clean}$ 
are fed into $Decoder_{clean}$. 
For COVID-19 CT images, the decomposed features are then aggregated again via 
the channel-wise summation operation and fed into $Decoder_{clean\&noisy}$ to 
reconstruct the volumetric airway structures.

\subsection{Refinement of Transferable Features}
As is mentioned before, the annotation of the public 
airway dataset is overall coarse and unsatisfactory. 
Since we have determined which features to transfer, 
then the transferable features can be further refined to be 
more robust through feature recalibration and 
introducing signed distance map.

\noindent{\bfseries Feature Recalibration:}
3D Channel SE(cSE) module~\cite{anatomynet} is designed to 
investigate the channel-wise attention. 
We embed this module between $Encoder_{clean}$ and $Decoder_{clean}$, 
aiming to refine the transferable features. 
Take $\mathbf{U}$ as 
input and $ \mathbf{\widetilde{U}}$ as output, $\mathbf{U},
\mathbf{\widetilde{U}} \in \mathbb{R}^{F \times D \times H \times W}$ with the number 
of channels {\itshape F}, depth {\itshape D}, height {\itshape H}, width {\itshape W}. 
3D cSE firstly compresses the spatial domain then 
obtains channel-wise dependencies $\mathbf{\widetilde{Z}}$, which are formulated as follows:
\begin{gather}
    Z_i = \frac{1}{D}\frac{1}{H}\frac{1}{W}\sum_{d=1}^{D}\sum_{h=1}^{H}\sum_{w=1}^{W}U_i(d,h,w),\quad i = 1,2,...,F \\
    \mathbf{\widetilde{Z}} = \sigma(\mathbf{W}_2\delta(\mathbf{W}_1\mathbf{Z})),
\end{gather}
where $\delta(\cdot)$ denotes 
the ReLU function and $\sigma(\cdot)$ 
refers to sigmoid activation, 
$\mathbf{W}_1 \in\mathbb{R}^{\frac{F}{r_c} \times F}$, 
and $\mathbf{W}_2 \in\mathbb{R}^{F \times \frac{F}{r_c}}.$ 
The $r_c$ represents the reduction factor in the channel domain, 
similar to~\cite{SE}. 
The output of 3D cSE is obtained by: 
$\mathbf{\widetilde{U}} = \mathbf{U} \odot \mathbf{\widetilde{Z}}$.

\noindent{\bfseries Signed Distance Map:}
In recent years, introducing the distance transformed map into CNNs have proven 
effectivity in medical image segmentation task\cite{karimi2019reducing,navarro2019shape,xue2020shape} 
due to its superiority of paying attention to the global structural information. 
The manual annotation of the plentiful tenuous bronchi is error-prone and often be labeled thinner or thicker. 
The 3D FCNs cooperating with common loss function treat the labeled foreground equally and intend to fit such 
coarse labels, which are difficult to extract robust features. 
Even though the thickness of the 
annotated bronchi is uncertain, the phenomenon of breakage or leakage in the 
annotation can be avoided by 
experienced radiologists. Therefore, the overall topology is correctly delineated, and we can use the 
topological structure instead of the coarse label as a supervised signal. Besides, the intra-class imbalance 
problem in airway segmentation is severe. Distance transform map is used to rebalance 
the distribution of trachea, main bronchi, lobar bronchi, and distal segmental bronchi. 
We use the signed distance map transform as a voxel-wise reweighting method, 
incorporating with the regression loss that focuses on the relatively small values 
(such as the lobar bronchi and distal segmental bronchi) by having larger gradient magnitudes.

Given the airway as target structure and each voxel $x$ in the volume set $X$, we 
construct the Signed Distance Map (SDM) function termed as $\bm{\phi}(x)$, defined as:
\begin{equation}
\label{eq5}
\bm{\phi}(x)=\left\{
\begin{aligned}
&0, & & x \in \rm {airway} \ \rm{and} \ x \in \mathcal{C},\\
-\inf\limits_{\forall z \in \mathcal{C}} &\left\|x-z\right\|_2, & & x \in \rm {airway} \ \rm{and} \ x \notin \mathcal{C},\\
+\inf\limits_{\forall z \in \mathcal{C}}& \left\|x-z\right\|_2, & & x \notin \rm {airway},
\end{aligned}
\right.
\end{equation}
where the $\mathcal{C}$ represents the surface of the airway, we further 
normalize the SDM into $[-1, +1]$. 
We then transformed the segmentation task on clean CT scans to an SDM regression problem and 
introduce the loss function that penalizes the prediction SDM for having the wrong sign and 
forces the 3D CNNs to learn more robust features that contain topological features for airway. 
Denote the $y_x$ as the ground truth of SDM and $f_x$ as the prediction of the SDM, 
the loss function for the regression problem can be defined as follows:
\begin{gather}
    L_{reg} = \sum\limits_{\forall x}\left\|f_x-y_x\right\|_{1} - \sum\limits_{\forall x}\frac{f_xy_x}{f_xy_x+f_x^2+y_x^2},
\end{gather}
where $\left\|\cdot\right\|_{1}$ denotes the $L_1$ norm.

\subsection{Training Loss Functions}
The training loss functions consist of two parts, 
The first part is the $L_{reg}$ for the clean CT scans, 
and the second part is the $L_{seg}$ for the 
noisy CT scans, we combine the Dice\cite{vnet} and Focal loss\cite{lin2017focal} to 
construct the $L_{seg}$:
\begin{gather}
    L_{seg} = -\frac{2\sum\nolimits_{\forall x}p_xg_x}{\sum\nolimits_{\forall x}(p_x+g_x)} -\frac{1}{|X|}(\sum\limits_{\forall x}(1-p_x)^2log(p_x)),
\end{gather}
where $g_x$ is the binary ground truth and $p_x$ is the prediction. The total loss is 
defined as $L_{total} = L_{seg} + L_{reg}$.

\section{Experiments and Results}
{\bfseries Dataset:} We used two datasets to evaluate our method. 
\begin{itemize}
    \item Clean Domain: Binary Airway Segmentation (BAS) dataset\cite{qin2019airwaynet}. It contains 90 CT 
    scans (70 CT scans from LIDC\cite{armato2011lung}) and 20 CT scans from the training set 
    of the EXACT'09 dataset\cite{lo2012extraction}. The spatial resolution 
    ranges from 0.5 to 0.82 mm and the slice thickness ranges from 0.5 to 1.0 mm. We randomly 
    split the 90 CT scans into the training set (50 scans), validation set (20 scans), and 
    test set (20 scans). 
    \item Noisy Domain: COVID-19 dataset. We collected 58 COVID-19 patients from three hospitals and 
    the airway ground truth of each COVID-19 CT scan was corrected by three 
    experienced radiologists. The spatial resolution of the 
    COVID-19 dataset ranges from 0.58 to 0.84 mm and slice thickness varies from 0.5 to 1.0 mm. 
    The COVID-19 dataset is randomly divided into 10 scans for training and 48 scans for testing.
\end{itemize}

\noindent{\bfseries Network Configuration and Implementation Details:} 
As shown in Fig.\ref{fig2}, each block in the encoder or decoder contains 
two convolutional layers followed by pReLU and 
Instance Normalization\cite{ulyanov2016instance}. 
The initial number of channel is set to 32, thus \{$F_1,F_2,F_3.F_4$\} = \{32,64,128,256\}. 
During the preprocessing procedure,
we clamped the voxel values to $[-1200,600]$ HU, normalized 
them into $[0,255]$, and cropped the lung field to remove unrelated background regions. 
We adopted a large input size of $128\times224\times304$ CT cubes 
densely cropped near airways and chose a batch size of 1 
(randomly chose a clean CT scan and noisy COVID-19 CT scan) in the training phase. 
On-the-fly data augmentation included the random horizontal flipping and random
rotation between $[-10^\circ,10^\circ]$. All models were trained 
by Adam optimizer with the initial learning rate of 0.002. The total 
epoch is set to 60 and 
the learning rate was divide by 10 in the $50^{th}$ epoch, the hyperparameter of 
$r_c$ used in 3D cSE module is set to 2. Preliminary 
experiments confirmed training procedures converged under this setup. 
In testing phase, we performed the sliding window prediction with stride 48. 
All the models were implemented in PyTorch framework with a single 
NVIDIA Geforce RTX 3090 GPU (24 GB graphical memory).

\noindent{\bfseries Evaluation Metrics:} We adopted three metrics to evaluate methods, 
including the a) tree length detected rate (Length)\cite{lo2012extraction}, 
b) branch detected rate (Branch)\cite{lo2012extraction}, and c) Dice score coefficient (DSC). 
All metrics are evaluated on the largest component of each airway segmentation result.

\begin{table}[t]
    \arrayrulecolor{gray}
    \caption{Results(\%) on the test set of the COVID-19 dataset. 
    Values are shown as mean $\pm$ standard deviation. `B' indicates the 
    training set of the BAS dataset and `C' indicates the training set of the COVID-19 dataset.}\label{airway_table1}
    \centering
    \begin{tabular}{@{}l|c|c|c|c@{}}
    \toprule
    \multicolumn{2}{c|}{Method}                                                                                               & Length              & Branch              & DSC                 \\ \hline
    \multicolumn{1}{c|}{\multirow{4}{*}{\begin{tabular}[c]{@{}p{1.4cm}<{\centering}@{}}3D\\UNet\end{tabular}}} & Train on B only            & 72.4 ± 4.8          & 62.1 ± 4.5          & 93.2 ± 1.5          \\ \cline{2-5} 
    \multicolumn{1}{c|}{}                                                                        & Train on B + C             & 82.8 ± 4.8          & 83.8 ± 3.8          & 95.2 ± 1.3          \\ \cline{2-5} 
    \multicolumn{1}{c|}{}                                                                        & Train on C only            & 85.7 ± 5.1          & 84.9 ± 3.5          & 95.9 ± 1.2          \\ \cline{2-5}
    \multicolumn{1}{c|}{}                                                                        & Train on B, finetuned on C & 86.8 ± 5.3          & 85.0 ± 4.1          & 95.7 ± 1.1          \\ \hline
    \multicolumn{2}{l|}{3D UNet + cSE (Medical Physics,2019)\cite{anatomynet}}                                                                 & 86.2 ± 5.3          & 84.6 ± 4.2          & 95.8 ± 1.2          \\ \hline
    \multicolumn{2}{l|}{Feature Alignment (TMI,2020)\cite{chen2020unsupervised}}                                                                         & 87.9 ± 4.9          & 85.5 ± 4.8          & 95.5 ± 1.6          \\ \hline
    \multicolumn{2}{l|}{Domain Adaptation (TPAMI,2018)\cite{rozantsev2018beyond}}                                                                       & 87.0 ± 4.6          & 84.9 ± 4.0          & 96.0 ± 1.3          \\ \hline
    \multicolumn{2}{l|}{Proposed w/o cSE\&SDM}                                                                                & 90.2 ± 5.3          & 87.6 ± 4.2          & 96.5 ± 1.2          \\ \hline
    \multicolumn{2}{l|}{Proposed w/o cSE}                                                                                     & 91.1 ± 4.3          & 86.8 ± 3.7          & \textbf{96.8 ± 1.0} \\ \hline
    \multicolumn{2}{l|}{Proposed w/o SDM}                                                                                     & 91.0 ± 4.7          & 86.3 ± 4.1          & 96.6 ± 1.0          \\ \hline
    \multicolumn{2}{l|}{Proposed}                                                                                             & \textbf{92.1 ± 4.3} & \textbf{87.8 ± 3.7} & \textbf{96.8 ± 1.1} \\ \bottomrule
    \end{tabular}
    \end{table}

\noindent{\bfseries Quantitative Results:} 
Experimental results showed that the way of training on the BAS dataset then 
evaluating on the COVID-19 dataset performed worst, as expected. 
Training merely on the COVID-19 dataset performed better than training on both 
the BAS dataset and the COVID-19 dataset, which 
implied the necessity of transfer learning 
rather than merely together different datasets. 
3D UNet with cSE \cite{anatomynet} was 
trained on the COVID-19 dataset and the results showed 
no significant improvement.
For comparison, three commonly used transfer learning methods, 
Fine-tuned (pre-trained on BAS dataset, fine-tuned on COVID-19 dataset), 
Feature Alignment (FA)\cite{chen2020unsupervised} through adversarial training, and 
Domain Adaptation (DA) by sharing weights\cite{rozantsev2018beyond}
were reimplemented to be applied in our task, the results in Table.\ref{airway_table1} 
demonstrated our proposed method is superior to these methods, 
the proposed method achieved the highest performance 
on all metrics of Length (92.1\%), Branch (87.8\%), and DSC (96.8\%).  
We also conducted the ablation study to investigate the effectiveness of each component of the 
proposed method. In Table \ref{airway_table1}, 
we observed that the original dual-stream network had 
outperformed the other methods, 
with the achievement of 90.2\% Length, 87.6\% Branch, and 96.5\% DSC. 
The improvement confirmed the 
validity of our proposed dual-stream network. Furthermore, 
cSE module and SDM could boost performance independently 
and the combination of cSE and SDM brings the highest performance 
gain, which demonstrated the necessity of refinement for transferable features.  
\begin{figure}[thbp]
    \centering
    \includegraphics[width=1.0\linewidth]{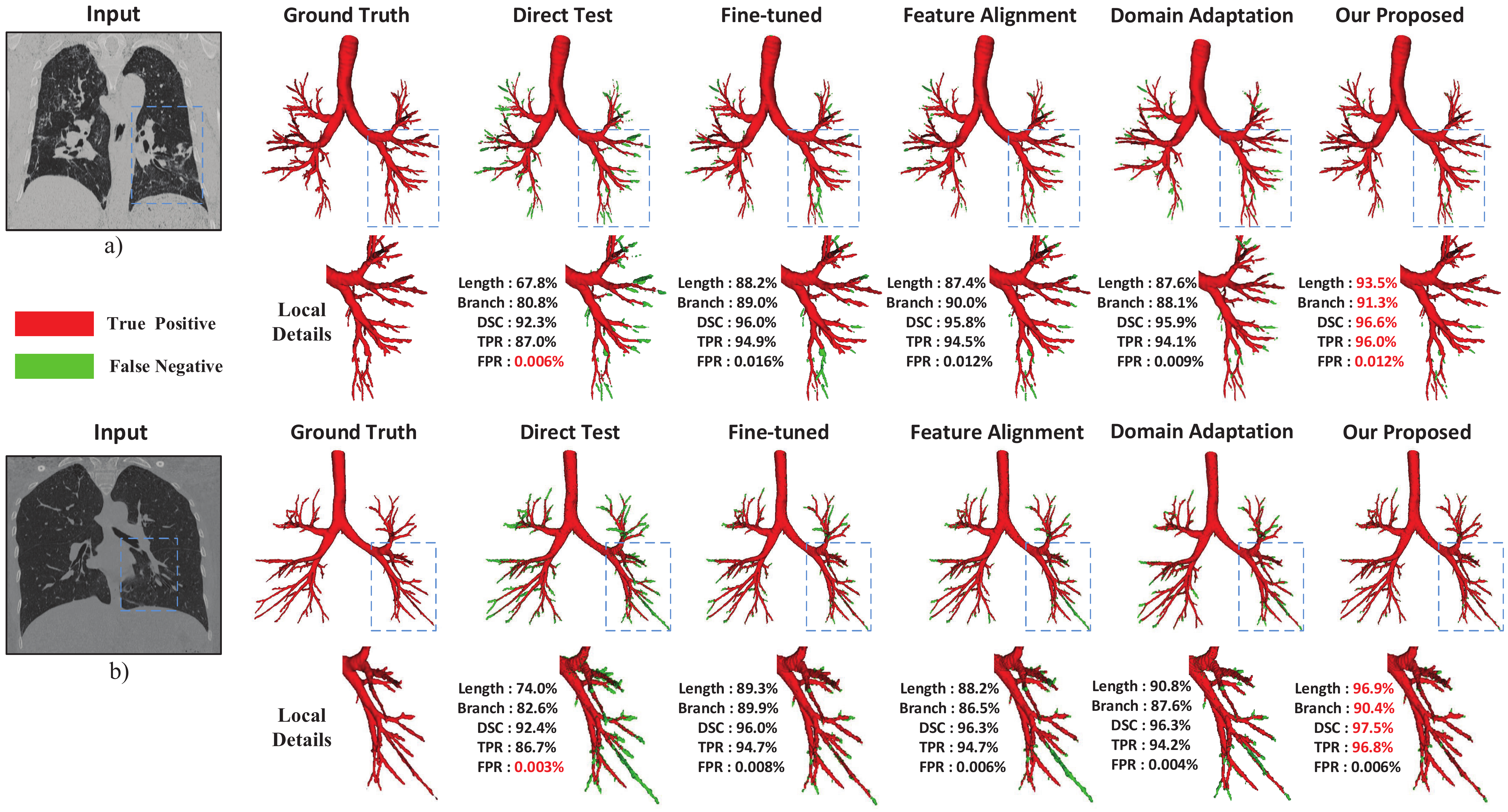}
    \caption{Visualization of segmentation results. a) is a severe 
    case and b) is a mild case in the test set of the COVID-19 dataset. The blue 
    dotted boxes indicate the local regions full of shadow patches and are 
    zoomed in for better observation.} \label{fig3}
\end{figure}

\noindent{\bfseries Qualitative Results:} 
The visualization of segmentation results is presented in Fig.\ref{fig3}. 
Compared to other methods, the proposed method gains 
improvement on both the severe and mild cases 
of the COVID-19 dataset, which accurately detected more bronchi surround by 
multifocal patchy shadowing of COVID-19.

\section{Conclusion}
This paper proposed a novel dual-stream network to learn transferable and robust 
features from clean CT scans to noisy CT for airway segmentation. Our proposed 
method not only extracted the transferable clean features but also extract 
unique noisy features separately, transferable features were further refined by 
the cSE module and SDM. Extensive experimental 
results showed our proposed method accurately segmented 
more bronchi than other methods.\\

\noindent{\bfseries Acknowledgements.} This work was partly supported by 
National Natural Science
Foundation of China (No. 62003208), 
National Key R$\&$D Program of China (No. 2019YFB1311503), 
Shanghai Sailing Program (No. 20YF1420800), Shanghai Municipal 
of Science and Technology Project 
(Grant No. 20JC1419500), 
and Science and Technology Commission of Shanghai Municipality under Grant 20DZ2220400.

%
%
%
\bibliography{paper8.bib}

\begin{thebibliography}{10}
\providecommand{\url}[1]{\texttt{#1}}
\providecommand{\urlprefix}{URL }
\providecommand{\doi}[1]{https://doi.org/#1}

\bibitem{armato2011lung}
Armato~III, S.G., McLennan, G., Bidaut, L., McNitt-Gray, M.F., Meyer, C.R.,
  Reeves, A.P., Zhao, B., Aberle, D.R., Henschke, C.I., Hoffman, E.A., et~al.:
  The lung image database consortium (lidc) and image database resource
  initiative (idri): a completed reference database of lung nodules on ct
  scans. Medical physics  \textbf{38}(2),  915--931 (2011)

\bibitem{chen2020unsupervised}
Chen, C., Dou, Q., Chen, H., Qin, J., Heng, P.A.: Unsupervised bidirectional
  cross-modality adaptation via deeply synergistic image and feature alignment
  for medical image segmentation. IEEE transactions on medical imaging
  \textbf{39}(7),  2494--2505 (2020)

\bibitem{chung2020ct}
Chung, M., Bernheim, A., Mei, X., Zhang, N., Huang, M., Zeng, X., Cui, J., Xu,
  W., Yang, Y., Fayad, Z.A., et~al.: Ct imaging features of 2019 novel
  coronavirus (2019-ncov). Radiology  \textbf{295}(1),  202--207 (2020)

\bibitem{SE}
Hu, J., Shen, L., Sun, G.: Squeeze-and-excitation networks. In: Proceedings of
  the IEEE conference on computer vision and pattern recognition. pp.
  7132--7141 (2018)

\bibitem{juarez2018automatic}
Juarez, A.G.U., Tiddens, H.A., de~Bruijne, M.: Automatic airway segmentation in
  chest ct using convolutional neural networks. In: Image Analysis for Moving
  Organ, Breast, and Thoracic Images, pp. 238--250. Springer (2018)

\bibitem{karimi2019reducing}
Karimi, D., Salcudean, S.E.: Reducing the hausdorff distance in medical image
  segmentation with convolutional neural networks. IEEE Transactions on medical
  imaging  \textbf{39}(2),  499--513 (2019)

\bibitem{lin2017focal}
Lin, T.Y., Goyal, P., Girshick, R., He, K., Doll{\'a}r, P.: Focal loss for
  dense object detection. In: Proceedings of the IEEE international conference
  on computer vision. pp. 2980--2988 (2017)

\bibitem{lo2012extraction}
Lo, P., Van~Ginneken, B., Reinhardt, J.M., Yavarna, T., De~Jong, P.A., Irving,
  B., Fetita, C., Ortner, M., Pinho, R., Sijbers, J., et~al.: Extraction of
  airways from ct (exact'09). IEEE Transactions on Medical Imaging
  \textbf{31}(11),  2093--2107 (2012)

\bibitem{vnet}
Milletari, F., Navab, N., Ahmadi, S.A.: V-net: Fully convolutional neural
  networks for volumetric medical image segmentation. In: 2016 Fourth
  International Conference on 3D Vision (3DV). pp. 565--571. IEEE (2016)

\bibitem{navarro2019shape}
Navarro, F., Shit, S., Ezhov, I., Paetzold, J., Gafita, A., Peeken, J.C.,
  Combs, S.E., Menze, B.H.: Shape-aware complementary-task learning for
  multi-organ segmentation. In: International Workshop on Machine Learning in
  Medical Imaging. pp. 620--627. Springer (2019)

\bibitem{ouyang2020dual}
Ouyang, X., Huo, J., Xia, L., Shan, F., Liu, J., Mo, Z., Yan, F., Ding, Z.,
  Yang, Q., Song, B., et~al.: Dual-sampling attention network for diagnosis of
  covid-19 from community acquired pneumonia. IEEE Transactions on Medical
  Imaging  \textbf{39}(8),  2595--2605 (2020)

\bibitem{qin2019airwaynet}
Qin, Y., Chen, M., Zheng, H., Gu, Y., Shen, M., Yang, J., Huang, X., Zhu, Y.M.,
  Yang, G.Z.: Airwaynet: a voxel-connectivity aware approach for accurate
  airway segmentation using convolutional neural networks. In: International
  Conference on Medical Image Computing and Computer-Assisted Intervention. pp.
  212--220. Springer (2019)

\bibitem{rozantsev2018beyond}
Rozantsev, A., Salzmann, M., Fua, P.: Beyond sharing weights for deep domain
  adaptation. IEEE transactions on pattern analysis and machine intelligence
  \textbf{41}(4),  801--814 (2018)

\bibitem{sun2016deep}
Sun, B., Saenko, K.: Deep coral: Correlation alignment for deep domain
  adaptation. In: European conference on computer vision. pp. 443--450.
  Springer (2016)

\bibitem{ulyanov2016instance}
Ulyanov, D., Vedaldi, A., Lempitsky, V.: Instance normalization: The missing
  ingredient for fast stylization. arXiv preprint arXiv:1607.08022  (2016)

\bibitem{wang2020noise}
Wang, G., Liu, X., Li, C., Xu, Z., Ruan, J., Zhu, H., Meng, T., Li, K., Huang,
  N., Zhang, S.: A noise-robust framework for automatic segmentation of
  covid-19 pneumonia lesions from ct images. IEEE Transactions on Medical
  Imaging  \textbf{39}(8),  2653--2663 (2020)

\bibitem{wang2020prior}
Wang, J., Bao, Y., Wen, Y., Lu, H., Luo, H., Xiang, Y., Li, X., Liu, C., Qian,
  D.: Prior-attention residual learning for more discriminative covid-19
  screening in ct images. IEEE Transactions on Medical Imaging  \textbf{39}(8),
   2572--2583 (2020)

\bibitem{wang2020weakly}
Wang, X., Deng, X., Fu, Q., Zhou, Q., Feng, J., Ma, H., Liu, W., Zheng, C.: A
  weakly-supervised framework for covid-19 classification and lesion
  localization from chest ct. IEEE transactions on medical imaging
  \textbf{39}(8),  2615--2625 (2020)

\bibitem{xue2020shape}
Xue, Y., Tang, H., Qiao, Z., Gong, G., Yin, Y., Qian, Z., Huang, C., Fan, W.,
  Huang, X.: Shape-aware organ segmentation by predicting signed distance maps.
  In: Proceedings of the AAAI Conference on Artificial Intelligence. vol.~34,
  pp. 12565--12572 (2020)

\bibitem{anatomynet}
Zhu, W., Huang, Y., Zeng, L., Chen, X., Liu, Y., Qian, Z., Du, N., Fan, W.,
  Xie, X.: Anatomynet: Deep learning for fast and fully automated whole-volume
  segmentation of head and neck anatomy. Medical physics  \textbf{46}(2),
  576--589 (2019)

\end{thebibliography}
\end{document}